%% file: arxiv.tex
\documentclass[a4paper, oneside, twocolumn, notitlepage, 10pt]{extarticle_ecoc}
\usepackage{ecoc}
\usepackage{comment}
\usepackage{color}
\usepackage{subfigure}
\usepackage{enumerate}
\usepackage{enumitem}
\usepackage{stfloats}
\usepackage{psfrag}
\usepackage{setspace}
\usepackage{balance}
\usepackage{enumitem}
\usepackage{tikz}
\usepackage{pgfplots,amsfonts}
\usepackage{pgfplotstable}
\usepackage{amsmath,amssymb}
\usepackage{xcolor}
\usepackage{graphicx}
\usepackage{tikz}
\usepackage{pgfplots}
\usepackage{float}
\usepackage{balance}
\usepackage{setspace}
\usepackage{xspace}
\usepackage[normalem]{ulem}

\usepackage{dsfont,empheq}
\usepackage{multicol,multirow}

\usepackage{booktabs, cellspace, hhline}

\usetikzlibrary{shapes}
\usetikzlibrary{spy}
\usetikzlibrary{fit}
\usetikzlibrary{shapes.multipart}
\usetikzlibrary{positioning}
\pgfplotsset{compat=1.14}
\usepackage{wrapfig}
\usepackage[acronym,nomain]{glossaries}
\input{acronyms.tex}
\usepackage{arydshln}

\definecolor{yellow}{RGB}{250, 199, 100} 

\begin{document}
\selectlanguage{english}    


\title{Analytical SNR Prediction in Long-Haul Optical Transmission using General Dual-Polarization 4D Formats 
}


\author{
    Zhiwei Liang\textsuperscript{(1)}, Bin Chen\textsuperscript{(1)}, Yi Lei\textsuperscript{(1)},
    Gabriele Liga\textsuperscript{(2)} and Alex Alvarado\textsuperscript{(2)}
}

\maketitle                  


\begin{strip}
 \begin{author_descr}

   \textsuperscript{(1)}  School of Computer Science and Information Engineering, Hefei University of Technology, China
   
   * \textcolor{blue}{\uline{bin.chen@hfut.edu.cn}}

  \textsuperscript{(2)} Department of Electrical Engineering, Eindhoven University of Technology, The Netherlands
  
 \end{author_descr}
\end{strip}

\setstretch{1.097}


\begin{strip}
  \begin{ecoc_abstract}
Nonlinear interference models for dual-polarization 4D (DP-4D) modulation have only been used so far to predict signal-signal nonlinear interference.  
We show that including the signal-noise term in the prediction of the effective signal-to-noise ratio in long distance DP-4D transmission improves the accuracy by up to 0.2~dB.
 \copyright 2022 The Author(s) 

  \end{ecoc_abstract}
\end{strip}

\section{Introduction}
Nonlinear interference (NLI) modeling in optical fiber transmission is a key tool to analyze the performance of optical communication systems and to optimize modulation formats. Various analytical models for
nonlinear fibre propagation have been proposed in the literature \cite{6685826, Dar13, Carena:14}. 
Among these models the enhanced Gaussian noise (EGN) model enables an accurate estimatation of the NLI induced by polarization-multiplexed 2D (PM-2D) formats, where two identical 2D formats are used to transmit information independently over two orthogonal polarization modes.~However, PM-2D formats are only a subset of all the possible dual-polarization four-dimensional (DP-4D) modulation formats, e.g., geometrically-shaped 4D formats \cite{Kojima2017JLT,8718568}.

Multidimensional modulation formats have been considered as an effective approach to harvest shaping gains \cite{ForneyJSAC1984}, especially in nonlinear optical fiber channel \cite{DarISIT2014}.~{In order to fully explore the potential of DP-4D modulation formats in the nonlinear fiber channel, 4D NLI models have been introduced in \cite{9343735,2020Extending} as a tool to efficiently find a trade-off between linear and nonlinear shaping gains \cite{GabrieleOFC2022,ChenOFC2022}}.

Under the additive NLI noise assumption, the effective signal-to-noise ratio (the SNR after fiber propagation and the receiver digital signal processing including chromatic dispersion compensation and phase compensation) for multi-span systems can be approximated as 
\vspace{-0.6em}
\begin{equation}\label{SNR}
\vspace{-0.6em}
\begin{split}
    \text{SNR}_{\text{eff}} &\triangleq \frac{P}{N_s\sigma^2_{ASE}+\sigma^2_{ss}+\sigma^2_{sn}},
\end{split}
\vspace{-0.6em}
\end{equation}
where $P$ denotes the transmitted signal power per channel, $N_s$ is the number of spans.~The total noise power consists of three parts: i) amplified spontaneous emission (ASE) noise over one span denoted as $\sigma^2_{ASE}$, ii) signal-signal (S-S) NLI power denoted as $\sigma^2_{ss}$ and iii) signal-ASE noise (S-N) NLI power denoted as $\sigma^2_{sn}$. 

In previous works, 4D NLI models have been validated and used only in terms of $\sigma^2_{ss}$ prediction. The impact of $\sigma^2_{sn}$ in the total effective SNR was thus neglected for general DP-4D formats.   

In this work, by assessing the contribution of signal-ASE noise interaction in the total NLI power, we analytically study the effective SNR in multispan amplified optical fiber transmission systems using general DP-4D formats.
This study is validated via split-step Fourier method (SSFM) simulations using various DP-4D modulation formats.~Our results show that including the S-N term can reduce the estimation error of the effective SNR by 0.2~dB, which can be translated into a 4\% prediction accuracy improvement in terms of transmission reach.

\section{Improving the Accuracy for 4D NLI Model}

To improve the accuracy of the effective SNR prediction, we study 
the impact of signal-ASE interaction for optimized 4D modulation formats based on the NLI model, which is built
on the fact that the x- and y-polarization could be dependent of one another \cite{2020Extending}.

For dual-polarized signals over single-channel transmission, the signal-signal NLI power $\sigma^2_{ss}$ in Eq.~(\ref{SNR}) can be approximated as \cite [Eq. (1)]{7831073}
\vspace{-0.6em}
\begin{equation}\label{ss}
\vspace{-0.6em}
\begin{split}
    \sigma^2_{ss} \approx \eta_{ss}N_s^{1+\varepsilon}P^3,
\end{split}
\vspace{-0.6em}
\end{equation}
where 
$\varepsilon$ is a coherence factor for self channel interference which is a function of fiber link parameters (attenuation, dispersion, span length, etc) \cite [Eq.(40)]{6685826}.~The $\eta_{ss}$ 
denotes the signal-signal NLI power coefficient over one span.
Here we denote the accumulated signal-signal NLI power coefficient over  $N_s$ spans as $\eta_{ss}^{(N_s)}=\eta_{ss} N_s^{1+\varepsilon}$. For general DP-4D formats, the modulation-dependent coefficient $\eta_{ss}^{(N_s)}$ for multi-span system can be calculated using Eq.~(1) in \cite{9489833}.

As we discussed in the introduction, the ASE noise generated by erbium-doped fibre amplifier (EDFA) leads not only to an additive white Gaussian noise (AWGN) but also to a nonlinear interference that produced by ASE noise and transmitted signal interaction \cite{7637002}. 
Under the assumption of flat transmitted signal spectrum and same propagated signal and ASE noise bandwidth, the signal-ASE NLI power coefficient can be estimated as $\eta_{sn}=3\eta_{ss}$ \cite{Cartledge:17,7831073}. Thus, by following \cite [Eq. (8)]{Cartledge:17}, the NLI power of signal-ASE interaction for DP-4D modulation  can be  derived as 
        \vspace{-0.6em}
\begin{equation}\label{sn}
        \vspace{-0.6em}
    \sigma^2_{sn}=\xi\eta_{sn}\sigma^2_{ASE}P^2=3\xi{\eta_{ss}}\sigma^2_{ASE}P^2,
\end{equation}
where $\sigma^2_{ASE}$ is the power of ASE noise over one span, $\xi\approx\frac{N_s^{2+\varepsilon}}{2+\varepsilon}+\frac{N_s^{1+\varepsilon}}{2}$ is the signal-ASE NLI accumulation coefficient.

Therefore, by considering both  signal-signal and signal-ASE interaction, the NLI power can be estimated via  Eq. \eqref{ss} and  \eqref{sn}, where we can obtain $\eta_{ss}$ as $\eta_{ss}^{(N_s)}/N_{s}^{1+\varepsilon}$. 
Note that  $\eta_{ss}^{(N_s)}$ is a constant value (for a given system configuration) linked to the
contributions of both modulation-independent and modulation-dependent nonlinearities, thus NLI power is   also a function of the given 4D modulation format.

The optical system we consider in this work is a single channel, multi-span transmission system with a symbol rate of 45~GBaud and a root-raised-cosine filter with roll-off factor of 0.01\%. The fiber link has the following parameters: attenuation coefficient $\alpha = 0.2$~dB/km, dispersion parameter $\beta_2 = -21.7$~ps$^2$/km and nonlinear coefficient $\gamma = 1.3$~(W~km)$^{-1}$.
Each span consists of an 80~km single-mode fiber followed by an EDFA with a noise figure of 5~dB.

\begin{figure}[!tb]
 \vspace{-0.0em}
    \centering
       \centering
     {\includegraphics{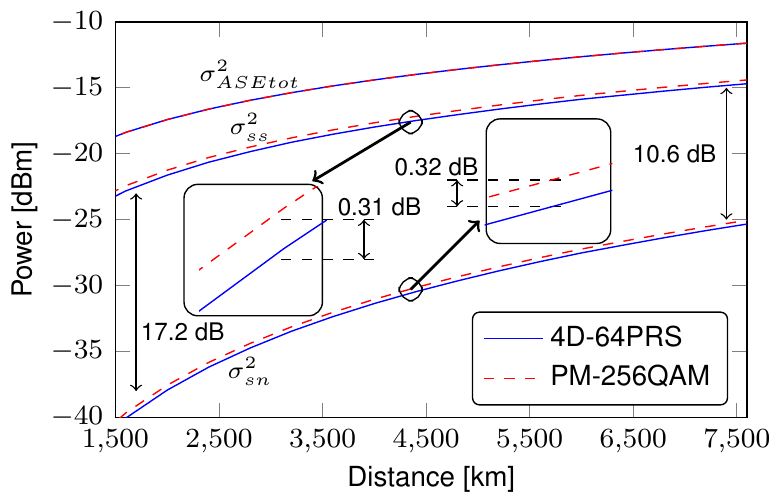}}
    \vspace{-1.2em}
    \caption{Noise power versus transmission distance at launch power of 0.5~dBm. Noise is shown separately, as  ASE noise, signal-signal NLI  and signal-ASE NLI.
    }
    \label{fig:P-sn}
\end{figure}

Fig.~\ref{fig:P-sn} shows the noise power, i.e., $\sigma^2_{ASEtot}=N_s\sigma^2_{ASE}$, $\sigma^2_{ss}$, $\sigma^2_{sn}$, against transmission distance.
Considering for example 4D-PRS64 at a distance of $1600$~km, $\sigma^2_{sn}$ differs from $\sigma^2_{ss}$ by a factor of 17.2~dB, while the difference is reduced to 10.6~dB for that of 7500~km.
 The proportion of $\sigma^2_{sn}$ in NLI power keeps increasing as the number of fiber span increases. 
 
To investigate the dependence of signal-ASE NLI on the modulation format, uniform square PM-256QAM is chosen as a baseline format and the NLI power is shown as dashed lines in Fig.~\ref{fig:P-sn}.~A 0.3~dB gap can be found when comparing these two modulation formats.
It is also shown that the gap between $\sigma^2_{sn}$ and $\sigma^2_{ss}$ decreases as the transmission distance increases. In particular, this gap reduces from 17.2 dB at 1,500 km to 10.6 dB at 7,500 km.~This indicates that the effect of signal-ASE NLI can not be fully neglected in very long-distance transmission. More results of modulation formats are shown in the next section.

\section{Simulation Results and Analysis}
{In this section, the accuracy of 4D model with S-S and S-N is validated via comparing with SSFM  for different 4D modulation formats. The SSFM simulates the nonlinear Manakov equation  with an uniform step size of 0.1~km.}

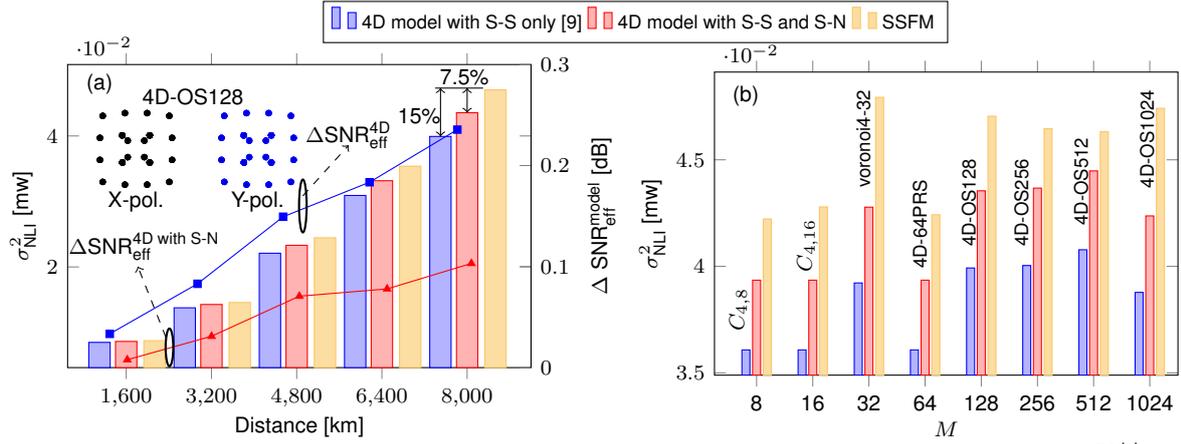
\begin{figure*}[!tbp]
 \vspace{-0.6em}
    \centering
      \centering
      \input{./tikz/fig2.tex}
     \vspace{-0.8em}
    \caption{Simulation results of multi-span optical fiber transmission with single channel: (a) NLI power and $\Delta \text{SNR}_{\text{eff}}^{\text{model}}$ vs. transmission distance for 4D-OS128 (inset); (b) NLI power for 4D various modulation formats at distance of 8000~km.}
    \label{fig:OS128}
\end{figure*}  

 In  Fig.~\ref{fig:OS128} (a) and (b), the estimation of NLI power are evaluated by using \romannumeral1) the 4D model with S-S only (blue bars)\footnote{Note that the 4D model is equivalent to EGN model for conventional PM-2D formats.}, \romannumeral2) 4D model with S-S and S-N (red bars), \romannumeral3)  SSFM (yellow bars) for different distances and modulation formats, respectively. 
 To target on a practical SD-FEC with 25\% overhead, 
 4D modulation formats are selected at required minimum SNR  in which $\text{GMI}=0.8m$~bit/4D for different spectral efficiencies with $m \in \{3,4,5,...,10\}$ from the existing 4D formats, which include 
the sphere packing database in \cite{Database}, some recently  proposed 4D  formats such as  4D-64PRS~\cite{8718568} and a family of 4D orthant-symmetric (OS)  formats \cite{ChenOFC2022}.\footnote{{The coordinates and labeling of these 4D modulation formats can be also found online at \href{https://github.com/TUe-ICTLab/Binary-Labeling-for-2D-and-4D-constellations}{https://github.com/TUe-ICTLab/Binary-Labeling-for-2D-and-4D-constellations}.}}

\begin{figure*}[!b]
\vspace{-1.2em}
\centering
{\includegraphics{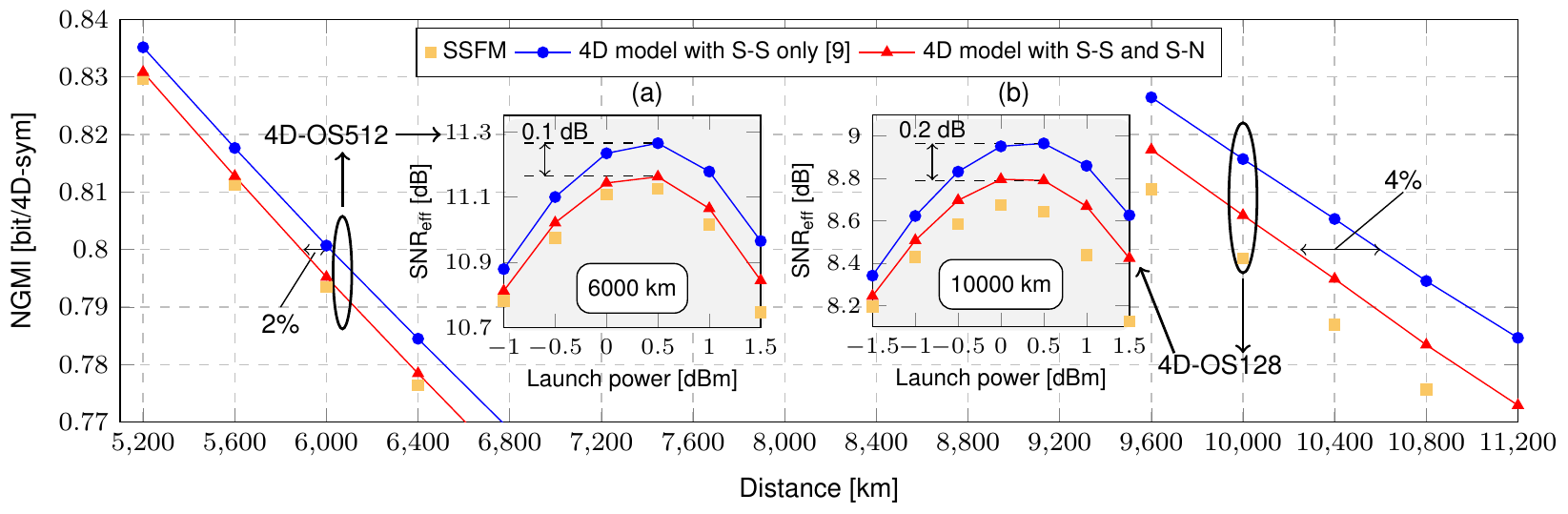}}
\vspace{-1.2em}
 \caption{ NGMI vs. transmission distance at optimal launch power for 4D-OS128 and 4D-OS512. Insets: $\text{SNR}_{\text{eff}}$ vs. launch power. }
\label{fig:ngmi}
\end{figure*}

Fig.~\ref{fig:OS128} (a) shows that the gap between our analytical predictions and SSFM becomes larger as distance increases for 4D-OS128 format \cite{BinChenJLT2021}. 
For a distance of 8000~km,  the 4D model with S-S underestimates NLI power by 15\% compared to SSFM, which can be halved by considering the S-N term.~{In order to translate this gap into effective SNR, we define the deviation of the effective SNR between a NLI model (4D or 4D model with S-N) estimation and the SSFM simulation as  $\Delta \text{SNR}_{\text{eff}}^{\text{model}} \triangleq \text{SNR}_{\text{eff}}^{\text{model}} - \text{SNR}_{\text{eff}}^{\text{SSFM}}$. For all distances shown, the deviation of 4D model with considering S-N interaction (red line in Fig.~\ref{fig:OS128} (a)) is within 0.1~dB.
}

Fig. \ref{fig:OS128} (b)  shows the NLI power estimation
for various modulation formats with different cardinalities $M$ over a distance of 8000~km. 
For all models shown, the tolerance of different 4D modulation formats to NLI is different. For example, the 4D-64PRS with constant modulus property has better nonlinear tolerance. In addition, for all 4D modulation formats shown, the 4D model with  S-N can improve the prediction accuracy of NLI power.

Fig.~\ref{fig:ngmi} shows the transmission performance estimation in terms of normalized generalized mutual information (NGMI) for the 4D models. It can be found that the 4D model with S-N can reduce the transmission reach prediction error by 2\% and 4\%, when compared to the 4D model with S-S only at NGMI of 0.8 for 4D-OS512 and 4D-OS128, respectively. The prediction accuracy gains come from reducing the 4D model overestimation of $\text{SNR}_\text{eff}$ compared to the 4D model with S-N. As shown in the insets (a) and (b) of Fig.~\ref{fig:ngmi}, the 4D model with S-N reduces the gap from SSFM by 0.1~dB at 6000~km for 4D-OS512 and by 0.2~dB at 10000~km for 4D-OS128 compared to accounting only for the S-S term.~Therefore, the 4D model with S-N could provide a better accuracy on  performance prediction  than 4D model, especially in long-distance transmission.

\section{Conclusion}
In this paper, we evaluated 
the weight of signal-ASE noise  interaction in the prediction of the effective SNR of general DP-4D constellations. Our results show that when signal-ASE noise interactions are considered the accuracy of SNR estimation is improved by 0.2~dB with respect to using existing 4D NLI models to compute only the signal-signal NLI contribution. 
~Providing an analytical expression for the signal-ASE noise interaction may improve the design of nonlinear-tolerant 4D modulation formats in long-haul systems.~Future work will focus on the design of DP-4D formats minimising the joint contribution of signal-signal and signal-noise NLI.

\vspace{0.1em}
\begin{spacing}{1}
{\footnotesize
\linespread{1} \textbf{Acknowledgements}: 
This work was partially supported by the National Natural Science Foundation of China (62171175, 62001151),  and funded by the EuroTechPostdoc programme and the European Research Council (754462 and 757791). 
}
\end{spacing}
\printbibliography
\end{document}

%% file: acronyms.tex
\newacronym{CUT}{CUT}{channel under test}

\newacronym{KK}{KK}{Kramers-Kronig}
\newacronym{CSPR}{CSPR}{carrier-to-signal power ratio}
\newacronym{KKRX}{KKRx}{Kramers-Kronig Receiver}
\newacronym{SSBI}{SSBI}{signal-signal beat interference}

\newacronym{GS}{GS}{geometric shaping}
\newacronym{PS}{PS}{probabilistic shaping}
\newacronym{DSP}{DSP}{digital signal processing}
\newacronym{MIMO}{MIMO}{multiple-input multiple-output}
\newacronym{TDE}{TDE}{time domain equalizer}
\newacronym{FDE}{FDE}{frequency domain equalizer}
\newacronym{LMS}{LMS}{least means square}
\newacronym{DDLMS}{DD-LMS}{decision directed least means square}
\newacronym{FFE}{FFE}{feed-forward equalizer}
\newacronym{FBE}{FBE}{feedback equalizer}
\newacronym{BPS}{BPS}{blind phase search}

\newacronym{SMF}{SMF}{single-mode fiber}
\newacronym[plural=SSMFs]{SSMF}{SSMF}{standard single-mode fiber}
\newacronym[plural=FMFs]{FMF}{FMF}{few-mode fiber}
\newacronym{FMF12}{FMF12}{12 mode FMF}
\newacronym{MMF}{MMF}{multi-mode fiber}
\newacronym{SI}{SI}{step index}
\newacronym{GI}{GI}{graded index}
\newacronym{DCF}{DCF}{dispersion compensated fiber}

\newacronym{SDM}{SDM}{space division multiplexing}
\newacronym{MDM}{MDM}{mode division multiplexed}
\newacronym{WDM}{WDM}{wavelength division multiplexing}
\newacronym{DWDM}{DWDM}{dense wavelength division multiplexing}
\newacronym{LP}{LP}{linear polarized}
\newacronym[plural=MMUXs,firstplural=mode multiplexers]{MMUX}{MMUX}{mode multiplexer}
\newacronym{PL}{PL}{photonic lantern}
\newacronym{3DWG}{3DWG}{3D-waveguide}
\newacronym{MDL}{MDL}{mode dependent loss}
\newacronym{DGD}{DGD}{differential group delay}
\newacronym{DMGD}{DMGD}{differential mode group delay}
\newacronym{QSM}{QSM}{quasi-single-mode}
\newacronym{GIMMF}{GI-MMF}{graded-index multi-mode fiber}

\newacronym{SSB}{SSB}{single side band}
\newacronym{QPSK}{QPSK}{quadrature phase shift keying}
\newacronym{QAM}{QAM}{quadrature amplitude modulation}
\newacronym{RRC}{RRC}{root-raised-cosine}
\newacronym{4D-64PRS}{4D-64PRS}{
four-dimensional 64-ary polarization-ring-switching}
\newacronym{DP}{DP}{dual-polarization}
\newacronym[\glslongpluralkey=states-of-polarization]{SOP}{SOP}{state-of-polarization}
\newacronym{PM}{PM}{polarization-multiplexed}

\newacronym{ECL}{ECL}{external cavity laser}
\newacronym{CW}{CW}{continuous wave}
\newacronym[plural=DFBs]{DFB}{DFB}{distributed feedback laser}

\newacronym[plural=DACs]{DAC}{DAC}{digital-to-analog converter}
\newacronym{ADC}{ADC}{analog-to-digital converter}
\newacronym{PRBS}{PRBS}{pseudo-random bit sequence}
\newacronym{LO}{LO}{local oscillator}
\newacronym{EDFA}{EDFA}{erbium-doped fiber amplifier}
\newacronym{MZM}{MZM}{Mach-Zehnder modulator}
\newacronym{DP-MZM}{DP-MZM}{dual-polarization Mach-Zehnder modulator}
\newacronym{ChUT}{ChUT}{channel under test}
\newacronym{WSS}{WSS}{wavelength selective switch}
\newacronym[plural=VOAs]{VOA}{VOA}{variable optical attenuator}
\newacronym[plural=PDCRXs]{PDCRX}{PDCRX}{polarization diverse coherent receiver}
\newacronym{DSO}{DSO}{digital storage oscilloscope}
\newacronym{ASE}{ASE}{amplified spontaneous emission}
\newacronym{PBS}{PBS}{polarization beam splitter}
\newacronym{PD}{PD}{photodiode}
\newacronym{AOM}{AOM}{acousto-optic modulator}
\newacronym{BPD}{BPD}{balanced photo-diode}
\newacronym{OMFT}{OMFT}{optical multi-format transmitter}
\newacronym{DPIQ}{DP-IQM}{dual-polarization IQ-modulator}
\newacronym{ABC}{ABC}{automatic bias controller}
\newacronym{OTF}{OTF}{optical tunable filter}
\newacronym{LSPS}{LSPS}{loop-synchronous polarization scrambler}
\newacronym{OSA}{OSA}{optical spectrum analyzer}

\newacronym{OSNR}{OSNR}{optical signal to noise ratio}
\newacronym{BER}{BER}{bit error rate}
\newacronym{IL}{IL}{insertion loss}

\newacronym{SDFEC}{SD-FEC}{soft-decision forward error correction}
\newacronym{HDFEC}{HD-FEC}{hard-decision forward error correction}
\newacronym{FEC}{FEC}{forward error correction}
\newacronym{LDPC}{LDPC}{low-density parity-check code}

\newacronym{AIR}{AIR}{achievable information rate}
\newacronym{AR}{AR}{achievable rates}
\newacronym{MI}{MI}{mutual information}
\newacronym{GMI}{GMI}{generalized mutual information}
\newacronym{BICM}{BICM}{bit-interleaved coded modulation}
\newacronym{GN}{GN}{Gaussian noise}

\newacronym{OVNA}{OVNA}{optical vector network analyzer}
\newacronym{NIR}{NIR}{near infrared}
\newacronym{CD}{CD}{chromatic dispersion}
\newacronym{OTDR}{OTDR}{optical time domain reflectometry}
\newacronym{OFDR}{OFDR}{optical frequency domain reflectometry}
\newacronym{GPU}{GPU}{graphics processing unit}
\newacronym{SVD}{SVD}{singular value decomposition}
\newacronym{WGN}{WGN}{white Gaussian noise}
\newacronym{AWGN}{AWGN}{additive white Gaussian noise}
\newacronym{PDL}{PDL}{polarization dependent loss}
\newacronym{SPS}{sps}{samples-per-symbol}
\newacronym{SE}{SE}{spectral efficiency}

%% file: tikz/fig2.tex
\hspace{-1em}
\subfigure{
\includegraphics{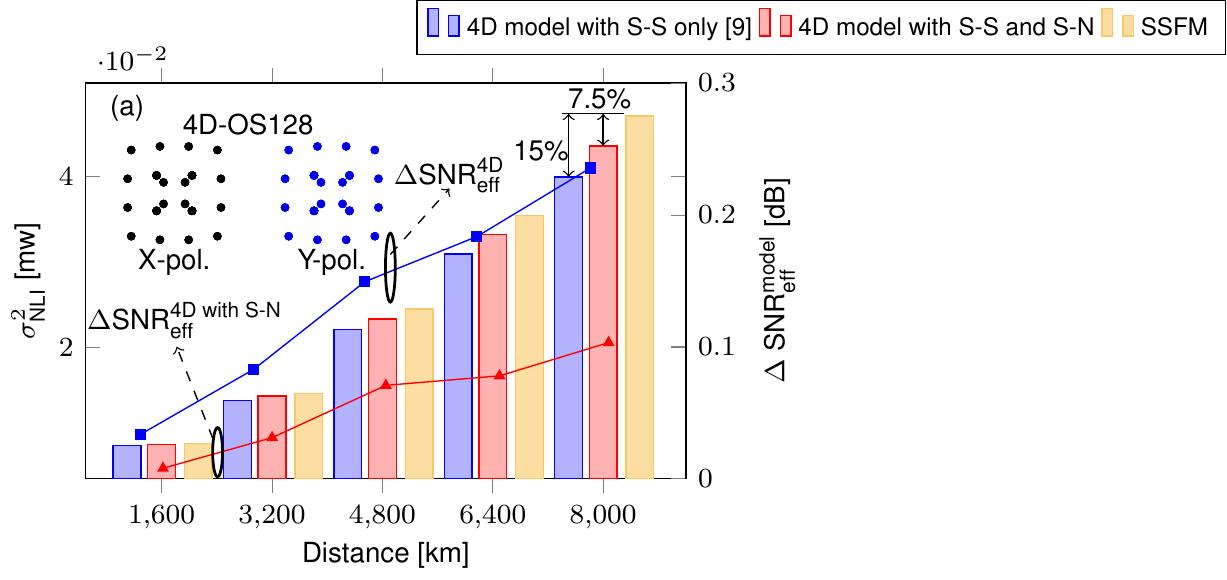}
}
\hspace{-13em}
\subfigure{
\includegraphics{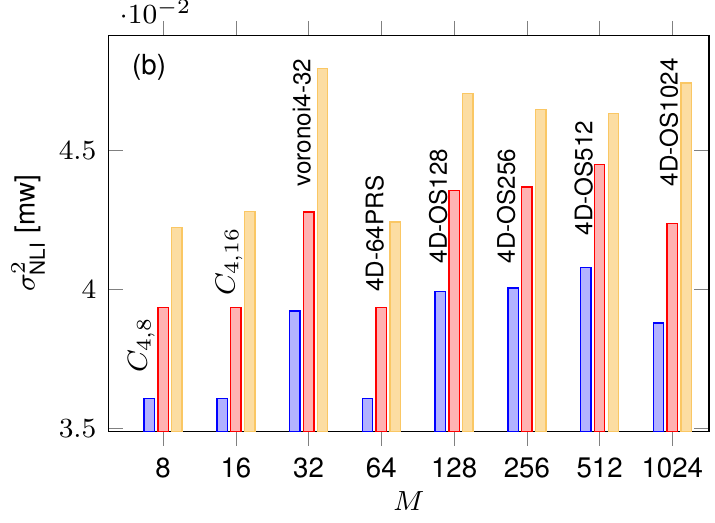}
}